\documentclass[twocolumn,apj]{emulateapj}
\usepackage{mathptmx,graphicx,grffile}

\usepackage[colorinlistoftodos, textwidth=4cm, shadow]{todonotes}
\usepackage{amsmath}
\usepackage{wasysym} 

\shorttitle{Density Evolution of Returning Filament Blobs}
\shortauthors{Carlyle et al.}

\begin{document}

\title{Investigating the Dynamics and Density Evolution of Returning Plasma Blobs from the 2011~June~7 Eruption}

\author{Jack~Carlyle\altaffilmark{1,2}, David~R.~Williams\altaffilmark{1}, Lidia~van~Driel-Gesztelyi\altaffilmark{1,3,4}, Davina~Innes\altaffilmark{2}, Andrew~Hillier\altaffilmark{5}, Sarah~Matthews\altaffilmark{1}}

\altaffiltext{1}{Mullard~Space~Science~Laboratory, University~College~London, Holmbury~St~Mary, Surrey, RH5~6NT, United~Kingdom}
\altaffiltext{2}{Max-Planck Institut f\"ur Sonnensystemforschung, 37191 Katlenburg-Lindau, Germany}
\altaffiltext{3}{LESIA-Observatoire de Paris, CNRS, UPMC Univ. Paris 06, Univ.
Paris-Diderot, 92195 Meudon, France}
\altaffiltext{4}{Konkoly Observatory, Budapest, Hungary}
\altaffiltext{5}{Kwasan and Hida Observatories, Kyoto University, Kyoto, Japan}
\email{j.carlyle@ucl.ac.uk}

\begin{abstract} 

This work examines infalling matter following an enormous Coronal Mass Ejection (CME) on 2011~June~7. The material formed discrete concentrations, or blobs, in the corona and fell back to the surface, appearing as dark clouds against the bright corona. In this work we examined the density and dynamic evolution of these blobs in order to formally assess the intriguing morphology displayed throughout their descent. The blobs were studied in five wavelengths (94, 131, 171, 193 and 211~\AA) using the Solar Dynamics Observatory Atmospheric Imaging Assembly (SDO/AIA), comparing background emission to attenuated emission as a function of wavelength to calculate column densities across the descent of four separate blobs. We found the material to have a column density of hydrogen of approximately 2~$\times$~10$^{19}$~cm$^{-2}$, which is comparable with typical pre-eruption filament column densities. Repeated splitting of the returning material is seen in a manner consistent with the Rayleigh-Taylor instability. Furthermore, the observed distribution of density and its evolution are also a signature of this instability. By approximating the three-dimensional geometry (with data from STEREO-A), volumetric densities were found to be approximately 2~$\times$~10$^{-14}$~g~cm$^{-3}$, and this, along with observed dominant length-scales of the instability, was used to infer a magnetic field of the order 1~G associated with the descending blobs.

\end{abstract}

\keywords{Sun: activity --- Sun: atmosphere --- Sun: coronal mass ejections (CMEs) --- Sun: filaments, prominences --- Sun: magnetic fields --- Sun: UV radiation}

\section{Introduction}

Eruptions occurring on the surface of the Sun cast huge amounts of matter and magnetic field out into the solar atmosphere and interplanetary space. These bundles of plasma with frozen-in magnetic fields can interact with the Earth's magnetosphere and have the potential to interfere with many different aspects of modern technology on large scales; for example, the geomagnetic storm which occurred in March 1989 caused the collapse of Hydro-Qu\'ebec's electricity transmission system, leaving 6 million people without electricity for 9 hours and costing the Canadian economy \$2 billion. For such reasons, `space weather' has become a focus of government interest in recent years, and a better fundamental understanding of the processes involved could lead to the ability to better predict events and protect our technologically-dependent world from this harsh, unforgiving environment.

These phenomena are closely coupled with the eruption of prominences -- condensations of chromospheric material suspended by magnetic forces in the corona. Since chromospheric material is relatively cool compared to the low-density corona, these structures appear dark when viewed on the solar disc (colloquially referred to as filaments), but are seen in emission in visible light when protruding off-limb (prominences) (see \citealp{Chen:2011wn} and references therein for a detailed description of these structures). The magnetic arcades and flux ropes thought to support prominences may become unstable, causing the material to rise and, ultimately, erupt, often resulting in what is known as a CME. Not all of the prominence material, however, will necessarily escape the Sun, with the magnetic field reconfiguring beneath the eruption and a new prominence often forming (perhaps a portion of flux-rope remains).

\begin{figure}
\centering
\includegraphics[scale=0.3]{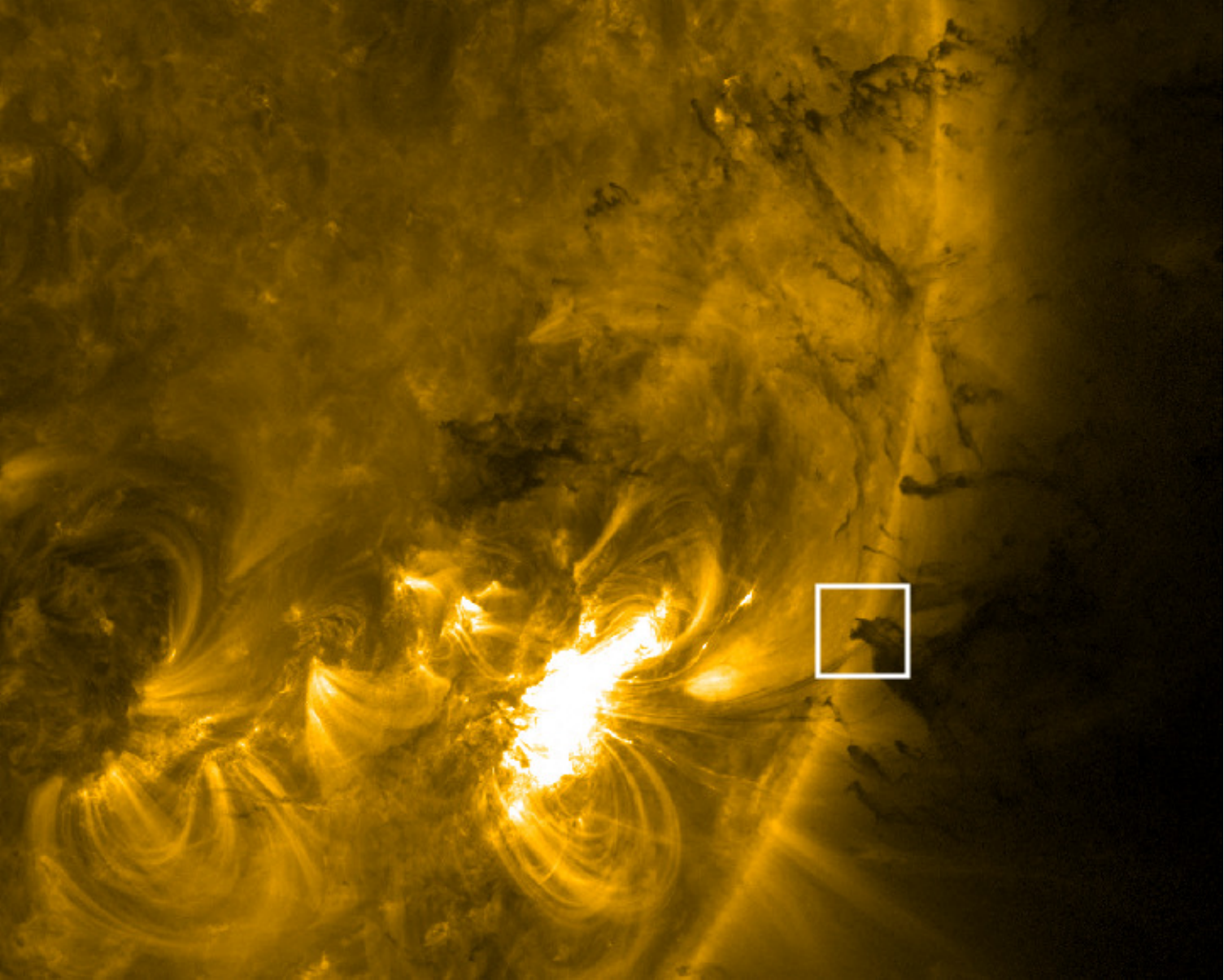}
\caption{Image of in-falling material as it crosses onto the solar disc at 07:05 UT - the morphology in the highlighted square may be compared with that of the material indicated in Figure~\ref{crab}. Highlighted blob is shown from STEREO in Figure~\ref{stereo} and is analysed in Figure~\ref{b1}.}
\label{snapb}
\end{figure}

On 2011~June~7, an immense CME took place; the active region precursor filament near the west solar limb rose and erupted, hurling an enormous amount of material into the solar atmosphere. While the filament itself did not appear particularly unusual prior to eruption, the sheer volume of ejected material seen returning to the solar surface is unparalleled in modern observation. The huge lateral expansion of the filament during the eruption is extremely eye-catching -- the vast area of the ejecta appears at least an order of magnitude larger than the initial foot-point separation, and suggests the filament carried a very large amount of mass. The material, which emerged as a single large cloud, appeared to repeatedly undergo the Rayleigh-Taylor (RT) instability above the solar limb \citep{Innes:2012cu}, highlighting a difference in density between the ejecta and the surrounding corona. As this cloud expanded, some of the material from the flanks of the CME appeared to stop moving outward and started falling back toward the solar surface, fragmenting into discrete condensations of matter: ``blobs''. This in-falling material passes back over the solar disc, appearing in absorption in the Extreme Ultraviolet (EUV) wavelengths, which indicates a finite density of lower-temperature material (neutral or singly-charged hydrogen or helium), compared with the surrounding and background atmosphere. A snapshot of this fallback is shown in Figure~\ref{snapb}, with one blob highlighted.

\begin{figure}
\centering
\includegraphics[scale=0.35]{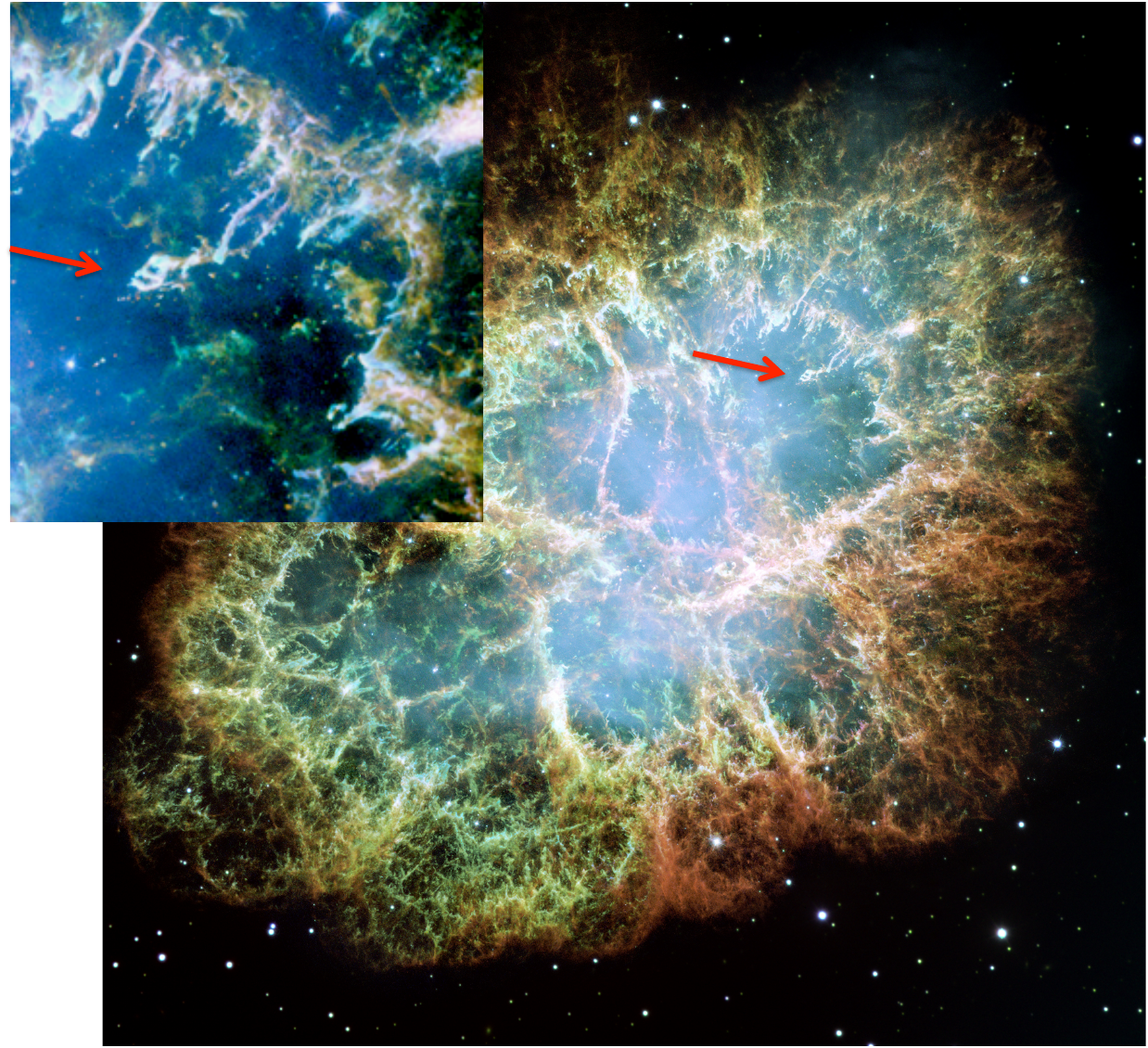}
\caption{The Rayleigh-Taylor instability as observed in the Crab Nebula. Credit: NASA, ESA and Allison Loll/Jeff Hester (Arizona State University). Acknowledgement: Davide De Martin (ESA/Hubble)}
\label{crab}
\end{figure}

Whilst blobs such as these have not been observed in the solar atmosphere until now (perhaps due to the lack of high-cadence observations rather than a lack of similar eruptions), these observed dynamics are not entirely unique. It is astonishing how morphologically similar the Crab Nebula appears in images captured by the Hubble Space Telescope, as can be seen in work by \cite{1996ApJ...456..225H}, which includes a rigorous study of these dynamics. This work concludes that the observed structures in this expanding supernova remnant are likely formed as a result of the RT instability, an image of which is presented in Figure~\ref{crab}. As such, it would be intriguing to investigate whether the observed blobs moving through the lower solar atmosphere undergo this instability as the higher material does \citep{Innes:2012cu}.

\begin{figure}
\centering
\includegraphics[scale=0.43]{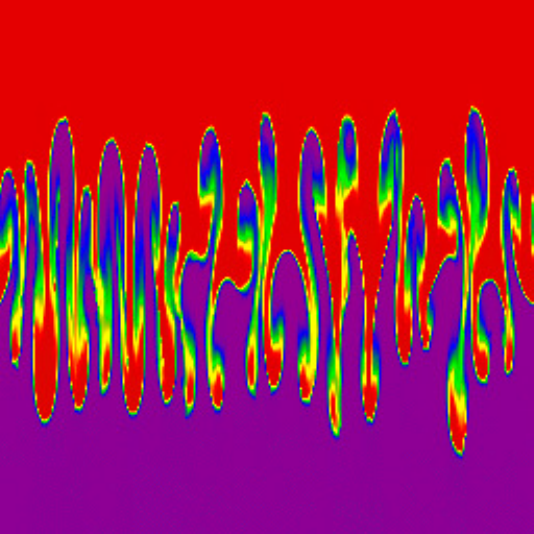}
\includegraphics[scale=0.43]{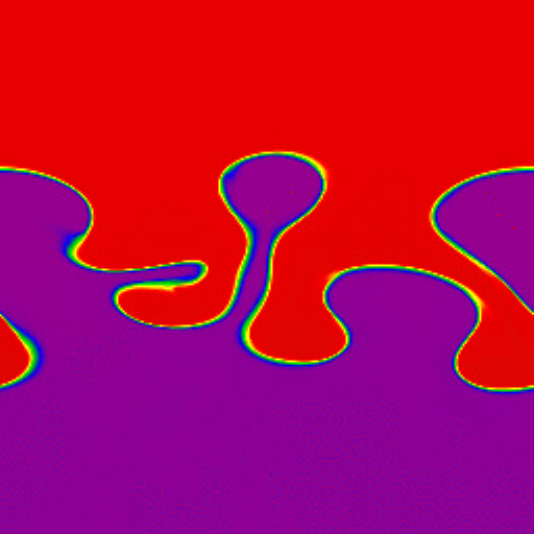}
\caption{Simulations of the Rayleigh-Taylor instability between materials of different density in the absence (left) and presence (right) of a magnetic field (parallel to the initially horizontal interface and in the plane of the image). A material of higher density lies above a material of lower density and is accelerated by gravity, demonstrating how a magnetic field modifies the morphology of the Rayleigh-Taylor instability. Generated by the Athena MHD Code.}
\label{mRT}
\end{figure}

In a nonmagnetic RT instability, the growth rate increases with decreasing scales, resulting in tightly bunched, long, thin fingers of high-density material progressing into low-density material, and vice versa. However, in the presence of a magnetic field there is a critical wavelength below which magnetic tension suppresses the instability, resulting in much wider fingers (with respect to their length) which also appear more cohesive. Figure~\ref{mRT} shows simulations of the RT instability under different magnetic constraints, generated by the Athena magnetohydrodynamic (MHD) code \citep{2008ApJS..178..137S}. This demonstrates how the undular mode of the magnetic Rayleigh-Taylor instability suppresses the formation of small-scale structure along the magnetic field.

A quantitative assessment of this instability may be made by considering its growth in the incompressible limit. Assuming that the magnetic field is purely in the $x$-direction (where $x$ is parallel to the interface between the two layers), the growth rate for such a perturbation is given as \citep{chandra61}:
\begin{equation} \label{rt1}
\gamma^2 = gkA - \frac{k^2 cos^2 \theta B^2_x}{2 \pi (\rho_u + \rho_l)}
\end{equation}
where g is the acceleration due to gravity and is perpendicular to $x$, $k$ is the wave number, A is the Atwood number (given as $(\rho_u - \rho_l)/(\rho_u + \rho_l)$), $\theta$ is the angle between $k$ and $B_x$, and $\rho$ is density ($u$ signifies the upper layer and $l$ signifies the lower layer). Taking the derivative with respect to k gives:
\begin{equation} \label{rt2}
2 \gamma \frac{\partial \gamma}{\partial k} = gA - \frac{2 k cos^2 \theta B^2_x}{2 \pi (\rho_u + \rho_l)}
\end{equation}
The most unstable growth rate will be at the peak of the distribution where $\partial \gamma / \partial k$ = 0, therefore the instability is described by:
\begin{equation} \label{rt2}
gA = \frac{2 k cos^2 \theta B^2_x}{2 \pi (\rho_u + \rho_l)}
\end{equation}
which rearranges to give:
\begin{equation} \label{domRT}
\frac{2 \pi}{k_u} = \lambda_u = \frac{2 cos^2 \theta B^2_x}{g(\rho_u - \rho_l)}
\end{equation}
where $\lambda_u$ is the dominant growth scale of the instability. It is then trivial to rearrange this equation to calculate $B_x$ from an observed growth scale:
\begin{equation} \label{domRT2}
B_x = \sqrt{\frac{g \lambda (\rho_u - \rho_l)}{2 cos^2 \theta}}
\end{equation}
These equations have been successfully applied to the Crab Nebula by \cite{1996ApJ...456..225H} and to prominence plumes on the Sun \citep{2010SoPh..267...75R, 2010ApJ...716.1288B}.

However, prominence mass (and therefore density, both before and after eruption) is not a trivial quantity to determine, as, historically, techniques have required spectroscopic observations in optically thick lines and radiative transfer modelling of these lines, leading to order-of magnitude estimates (see \citealp{Labrosse:2010bt} and references therein for a more detailed history). Using EUV imaging observations, \cite{Gilbert:2004vs} applied temporal-- and spatial--interpolative approaches to determine the column density of erupting and quiescent prominences, respectively, using absorption in the Fe XII (195) spectral band, and calculated the mass of an erupting prominence from 1999 July 12 to be approximately 6~$\times$~10$^{14}$~g.? \cite{Gilbert:2010hs} then expanded this technique to conduct the analysis in three different wavelength regimes, covering three different species' photoionisation continua. They concluded that the total prominence mass estimate is lower for the longer wavelengths analysed, attributed to the higher opacity in higher wavelengths causing a saturation of continuum absorption in these lines and thus a potentially large underestimation of the mass. This suggests that such column density diagnostics are best conducted at shorter wavelengths, where H$^0$, He$^0$ and He$^+$ are all ionised.

Analysing the falling prominence blobs in the 2011 June 7 filament eruption, \cite{2013ApJ...776L..12G} find that the brightenings observed when the blobs impacted the solar surface are likely due to compression of the plasma rather than reconfiguration of the local magnetic topology ({\it i.e.}, reconnection). \cite{Landi:2013bc} have also investigated the emission and absorption in this erupted material and find that the temperature is likely to be 33,100 $\pm$ 2,200 K with an electron density of $3.6^{+1.1}_{-0.7}~\times~10^{19}$~cm$^{-2}$. 

In this work, we analyse the density of these blobs and examine how this evolves as they fall through the solar atmosphere. We also examine the dynamics and distribution of mass in the blobs in order to determine whether the material undergoes the magnetic Rayleigh-Taylor instability. Finally, we use Equation~\ref{domRT2} to infer a magnetic field strength associated with the blobs under the instability.

\section{Observations}

This work uses images collected by the Solar Dynamics Observatory's Atmospheric Imaging Assembly (SDO/AIA: \citealp{2012SoPh..275...17L}) between 06:40 and 08:40 UT on 2011 June 7 in the 94, 131, 171, 193 and 211 \AA~ passbands. The eruption occurred from NOAA active region 11226, which was located in the vicinity of the south-west limb, and most of the in-falling material passed over this quadrant upon returning to the Sun (see the online content for a movie of the eruption and fall-back).

Each image in each wavelength was deconvolved using a Richardson-Lucy (RL) algorithm and wavelength-dependent point-spread functions as described by \cite{Grigis:2012wn}. This is to minimise the effect of diffraction and stray light in the images arising from the entrance filter and the focal plane filter on SDO/AIA.

\begin{figure}
\centering
\includegraphics[scale=0.45]{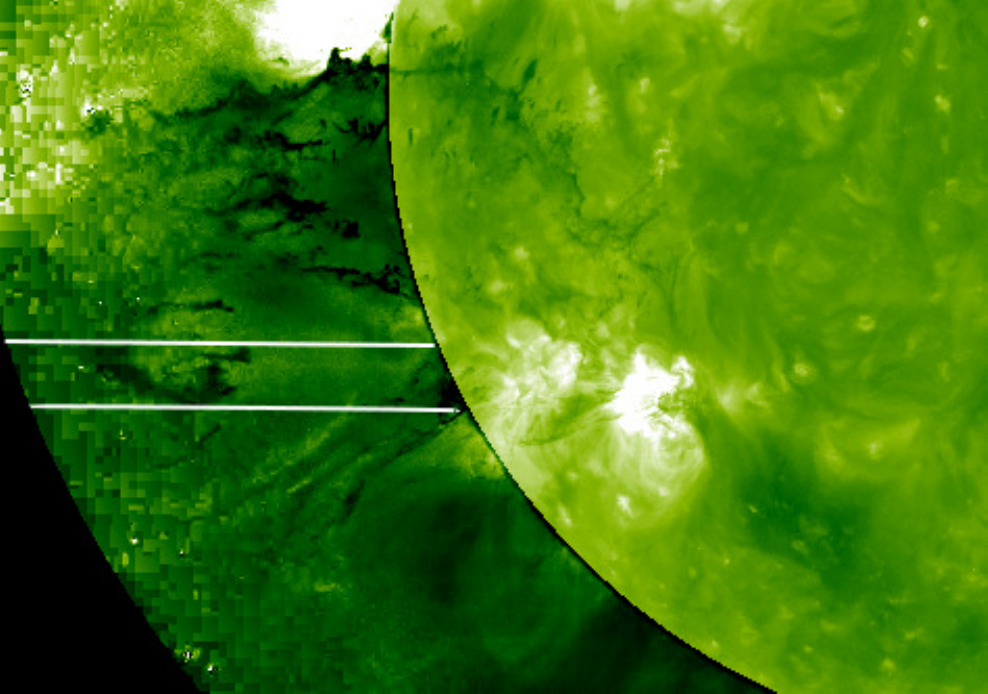}
\caption{The view of the in-falling material from STEREO-A at 07:05 UT. The white lines mark the line-of-sight of the top and bottom of the box shown in Figure~\ref{snapb}. Since there are only two views of the material, precise volume calculations cannot be performed; however the geometry of the material from two perspectives can aid estimates of the shapes and volumes of the blobs.}
\label{stereo}
\end{figure}

Also used were images collected by STEREO-A, which gives a different perspective on the eruption and a more complete idea of the blobs' geometry. A snapshot of the post-eruption fallback is shown in Figure~\ref{snapb} as seen from SDO, and from STEREO-A in Figure~\ref{stereo}. Note that the descent of these blobs has also been analysed by \cite{2013ApJ...776L..12G}.

The targets of our study were chosen based on a long, unobscured descent for each blob, in order to maximise the evolution timescale observed. The images used were taken at points in the descent where the blob appeared to lie above a relatively `quiet' region of the solar surface, in order to ensure that the image used as the background (co-spatial to the blob image) was as uniform as possible over time. The time steps were chosen to be at roughly equal intervals, although the requirement for `clean' background images also guided this choice. Four blobs were studied in total, with the blob exhibiting the most obvious apparent RT instability being studied at the highest cadence.

\section{Method}

In this study, we use the polychromatic opacity imaging method developed by \cite{Williams:2013bq} to obtain estimates of the column density of the erupted neutral hydrogen.? This technique works by measuring the absorption depth of the cooler material in five co-temporal AIA images, each at a different wavelength, using intensity measurements of the target image compared to a background image (a co-spatial image some minutes beforehand). The measured absorption depths, a function of wavelength, can then be fitted to a function of only two variables, one of which is column density.

Optical depth is related to density and intensity, respectively, using the following equations:
\begin{equation} \label{tau1}
\tau(\lambda)=N\sigma(\lambda)
\end{equation}
where $\tau$ is optical depth (a function of wavelength, $\lambda$), N is column density (particles per unit area) and $\sigma$ is cross-sectional area (also a function of wavelength); then, the observed intensity, $I$, is given by
\begin{equation} \label{intens1}
I=I_0e^{-\tau}
\end{equation}
where $I_0$ is the intensity that would be received in the absence of any obscuring material.
Equation \ref{intens1} may be re-written using observed intensity in the presence of the blob, $I_{obs}$ (given by the blob image), background intensity, $I_{b}$ (the intensity from material obscured by the blob), foreground intensity, $I_{f}$ (the emitted intensity from all material between the blob and the observer) and the unattenuated intensity, $I_{0}$ (given by the unocculted image, which is approximately the sum of the background and foreground intensity) and $f$, the pixel-filling factor:
\begin{equation} \label{intens2}
I_{obs}=I_b[fe^{-\tau}+(1-f)]+I_f 
\end{equation}
Rearranging, we are left with the following equation:
\begin{equation} \label{intens3}
1- \frac{I_{obs}}{I_0} =f\frac{I_b}{I_0}(1-e^{-\tau}))
\end{equation}
or
\begin{equation} \label{dg}
d(\lambda)=G(1-e^{-\tau(N,\lambda)})
\end{equation}
where the left-hand side of Equation \ref{intens3} is the absorption depth, $d(\lambda) = 1 - ({I_{obs}} \slash {I_0})$, and the geometric depth $G = f({I_{b}} \slash {I_0})$.

\begin{figure}
\centering
\includegraphics[width=0.45\textwidth]{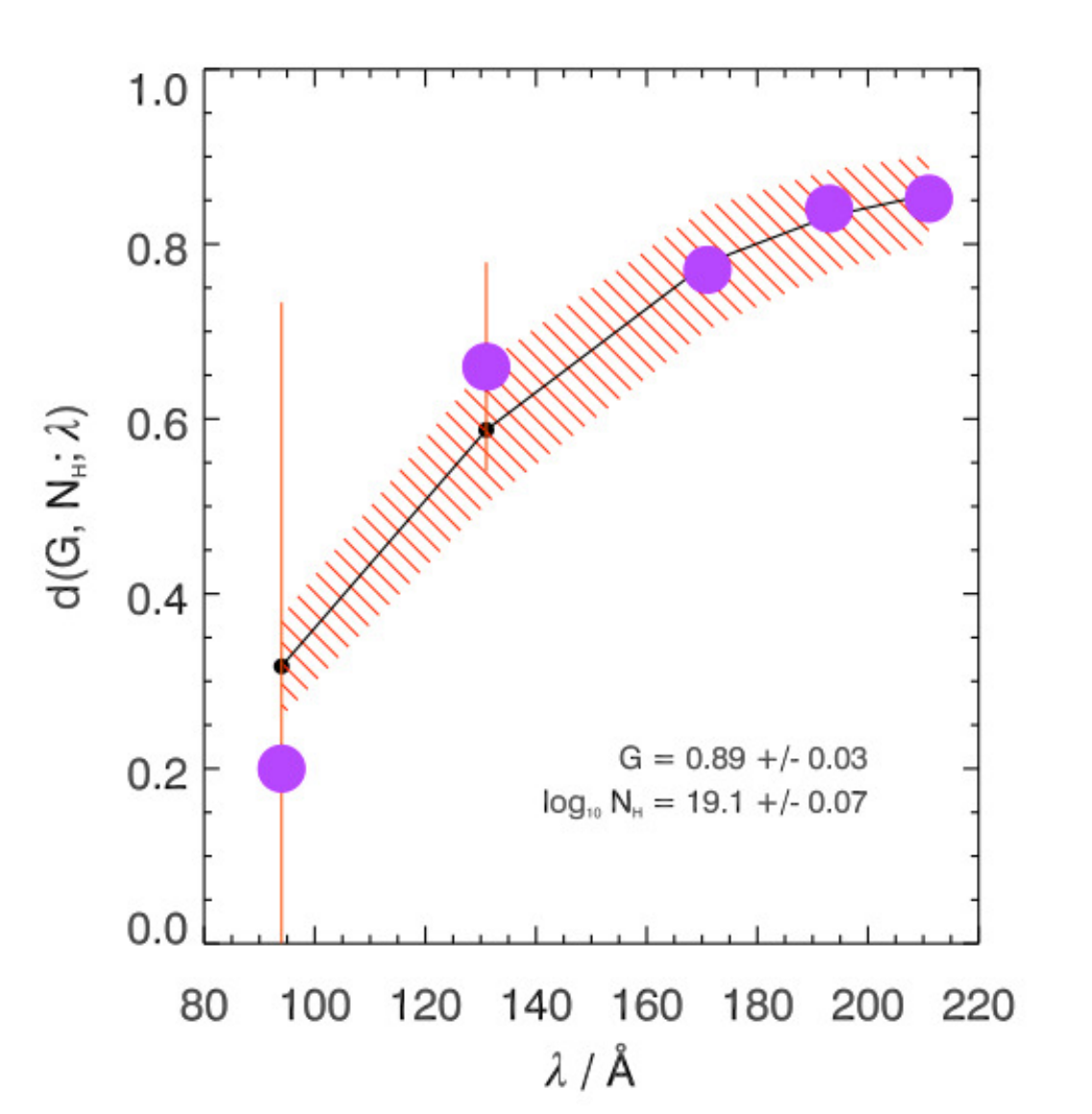}
\caption{This figure gives a graphical representation of Equation~\ref{dg} with the free parameters ($N_H$ and $G$) constrained such that the plot best lies over measured absorption depth as a function of wavelength (represented by the circles, calculated using the intensity ratio between the target and background).}
\label{davesgraph}
\end{figure}

The left-hand side of Equation \ref{dg} is observable, obtained by comparing the intensity in the background image with the image of the blob; the right-hand side is a model to which the calculated opacities from the five wavelengths may be fitted with only two variables, $G$ and $N_H$ (see Figure~\ref{davesgraph}). The value of $G$ is expected to be high across the whole blob, decreasing suddenly as the edge is reached as the pixel filling factor, $f$, tends to zero in the absence of absorbing material \cite{Williams:2013bq}, and a particular value of $G$ may be used as a definition of the edge of the blobs. The fitting can be done using a Levenberg-Marquardt least-squares minimisation algorithm, which returns a $\chi^2_\nu$ value which describes the accuracy of fit of the data to the model. A perfect fit would be indicated by a value of 1, with goodness-of-fit decreasing as $\chi^2_\nu$ moves further from unity in either direction.

The blobs studied were of various sizes, and maps of the column density were calculated at a number of points in time during the descent of each (generally 5 time-steps were used, though 15 were analysed for one particularly interesting blob). To examine how these structures evolved as they fell through the corona, and how well they correspond to the behaviour of the magnetic Rayleigh-Taylor (mRT) instability, the calculated values of column density in each blob were monitored throughout their descent (i.e. if a fluid is accelerated by gravity and encounters an interface, as is the case with the mRT instability, enhanced density is expected towards this interface as material ``piles-up''). Blobs were also compared with one another in order to ascertain whether they possess similar densities, which may indicate whether the physical conditions for the blob formation are due to similar density and/or mass values.

One blob was chosen based on morphology: it demonstrated repeated instances of the suspected instability, with large separation between consecutive fingers and a long, easy-to-follow descent through the solar atmosphere. The large separation suggests that these successive branches separate parallel to the plane of the sky (or close to); statistically, no orientation of magnetic field (parallel to the interface and perpendicular to the acceleration) should be favoured over another, and we expect scales in all blobs to be of the same order of magnitude (since the physical parameters should all reflect those of the progenitor cloud). It is therefore a trigonometric argument which states that the largest observed projected separations are likely to be (almost) in the plane of the sky \citep{1996ApJ...456..225H}. The depth of this blob was estimated to be of the same order of magnitude as the diameter using the STEREO-A data (shown in Figure~\ref{stereo}) and the column density was divided by this value to obtain a volumetric density estimate in order to formally assess the instability using Equation~\ref{domRT}.

\section{Results}

\begin{figure}
\centering
\includegraphics[width=0.5\textwidth,clip=,trim=80 40 100 20,angle=0]{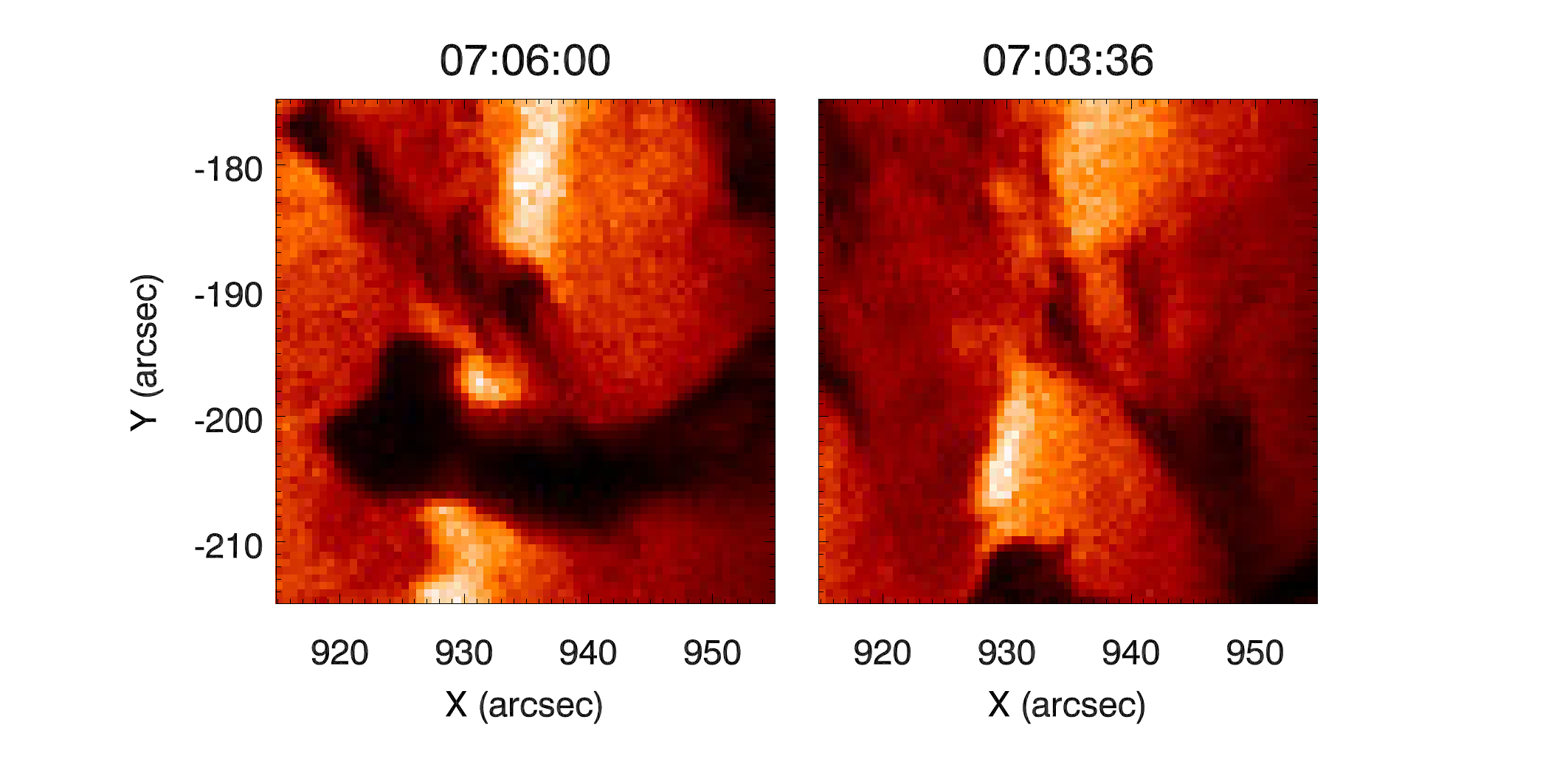}
\caption{171 \AA~images of a blob (at 07:06, left) and its associated background image (at 07:03, right). This blob is seen to move from the south-west of the image to the north-east.}
\label{171}
\end{figure}

\begin{figure}
\centering
\includegraphics[width=0.5\textwidth,clip=,trim=20 30 50 75,angle=0]{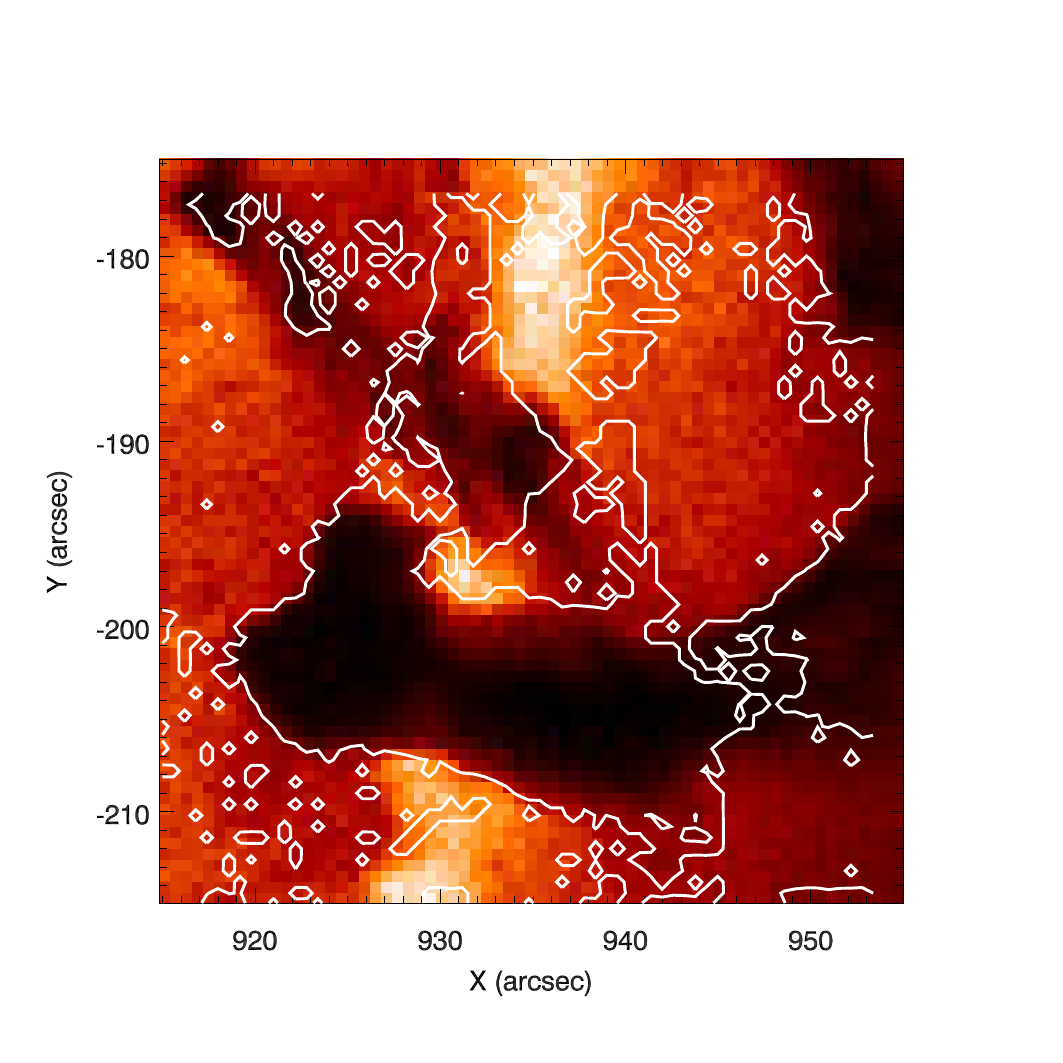}
\caption{171 \AA~images of blob shown in Figure~\ref{171} with contours of G = 0.5 overplotted.}
\label{171cont}
\end{figure}

\begin{figure}
\centering
\includegraphics[width=0.5\textwidth,clip=,trim=10 0 0 10,angle=0]{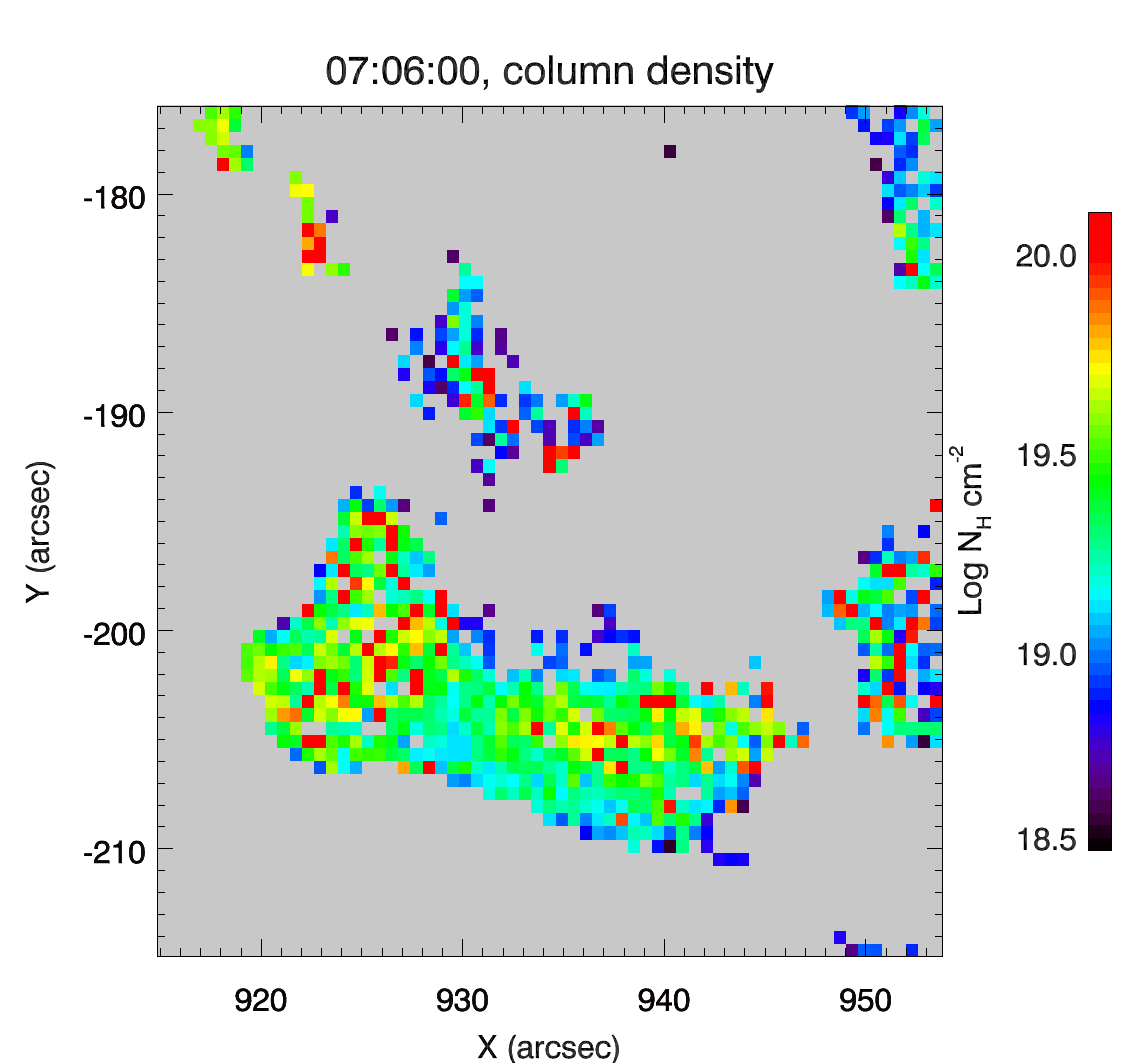}
\includegraphics[width=0.5\textwidth,clip=,trim=10 0 0 10,angle=0]{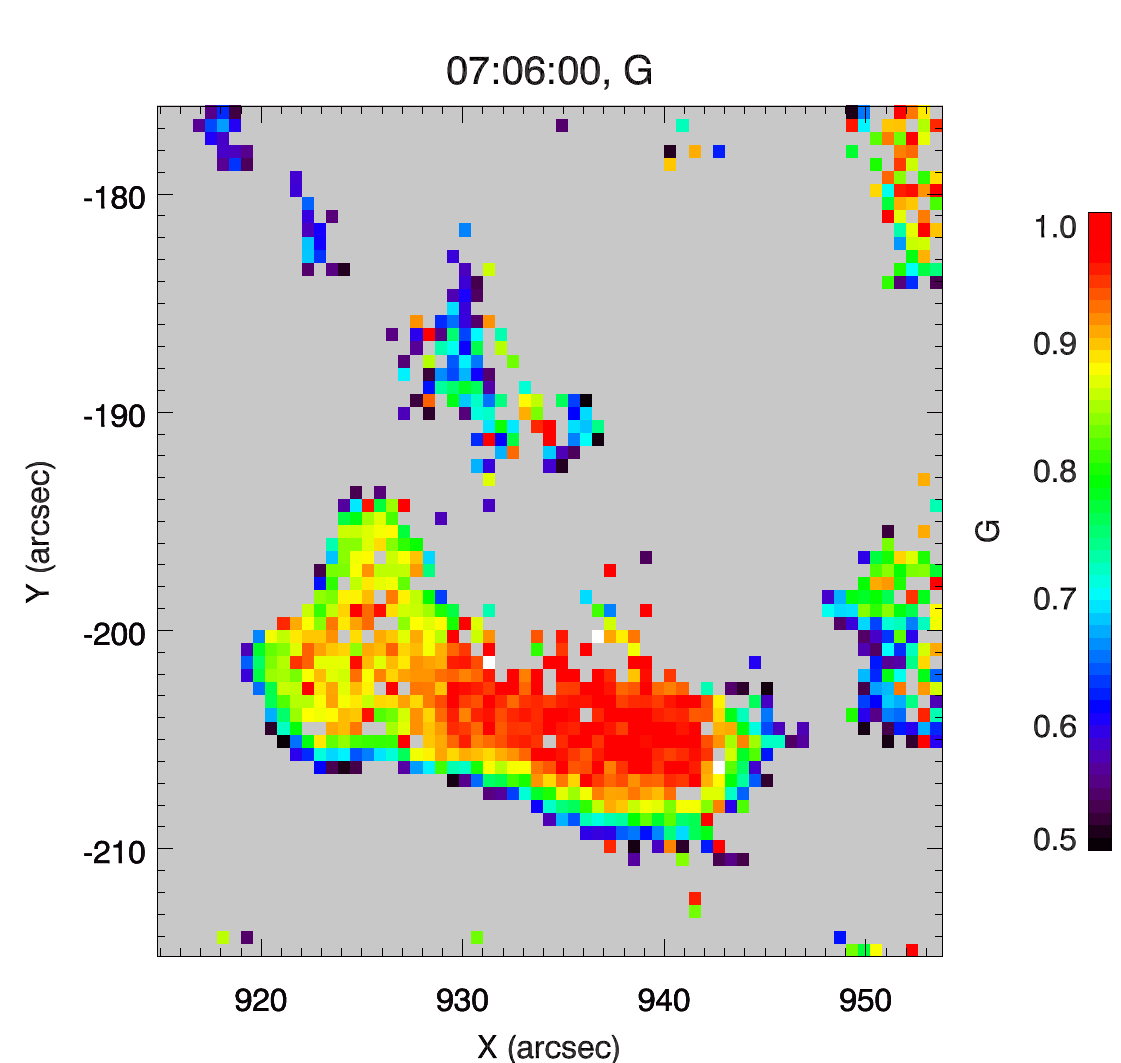}
\caption{Column density (top) and G (bottom) maps for the blob show in Figures~\ref{171}~and~{171cont}. Direction of travel is $\sim$40$^{\circ}$  to the negative X-axis.}
\label{A1}
\end{figure}

Figure~\ref{171} shows two images (in the 171 \AA~passband) used to determine column density -- the blob image and the background image (co-spatial, three minutes prior). 
Figure~\ref{171cont} shows the same blob image with contours of $G$ = 0.5 overlaid. 
$G$ is comprised of pixel filling factor, $f$, and fractional background emission, ${I_{b}} \slash {I_0}$, both of which are expected to be close to 1 in the pixels containing cool, dense matter high in the corona. This value can be described as the fraction of light per pixel that interacts with the blob material, and it can be said that when this value is greater than 0.5, the pixel is dominated by this material. Therefore a value of 0.5 was chosen to define the edge of the blobs, and in the figures showing $G$ and $N_H$ maps, grey pixels are all zones with a $G$ of less than 0.5 (or where $\chi^2_\nu$ is greater than 10 or less than 0.1). Figure~\ref{A1} shows the calculated column density and $G$ values for the same blob. 

\begin{figure*}
\centering
\includegraphics[height=4.5cm,trim=50 60 0 30,angle=0]{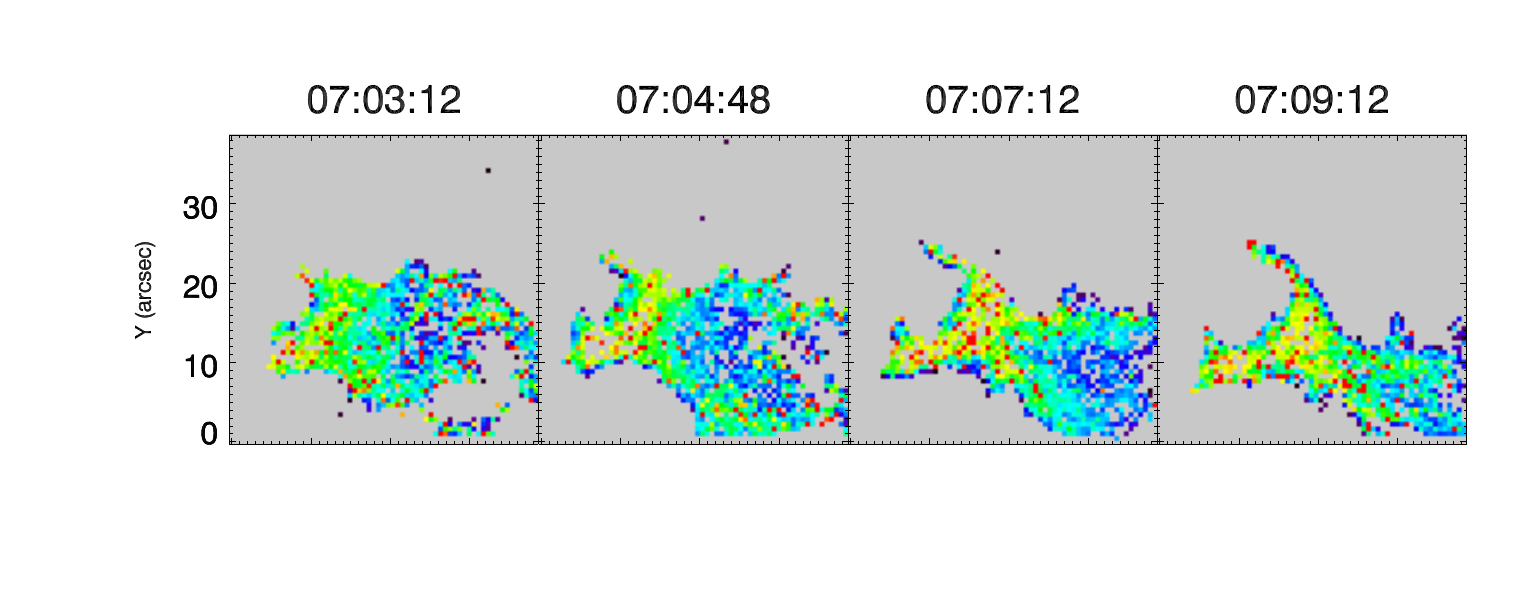} 
\includegraphics[height=4.5cm,trim=50 60 0 30,angle=0]{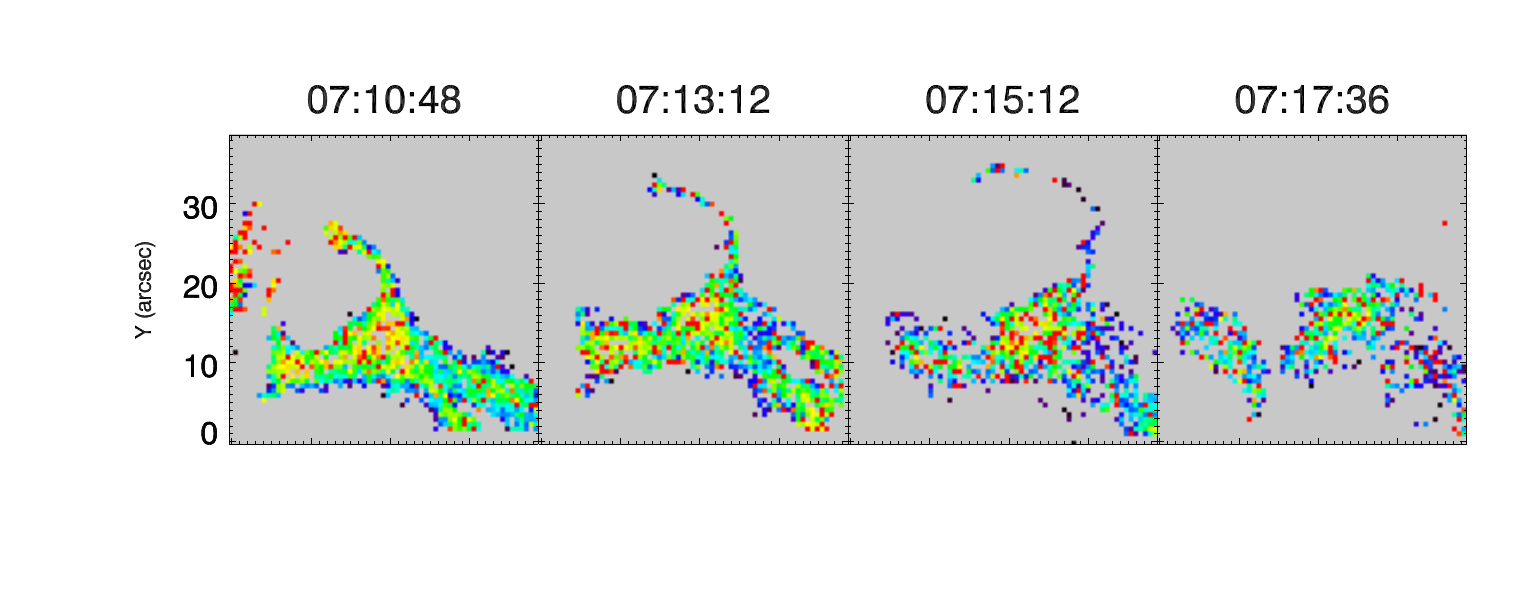} 
\includegraphics[height=4.5cm,trim=50 60 0 30,angle=0]{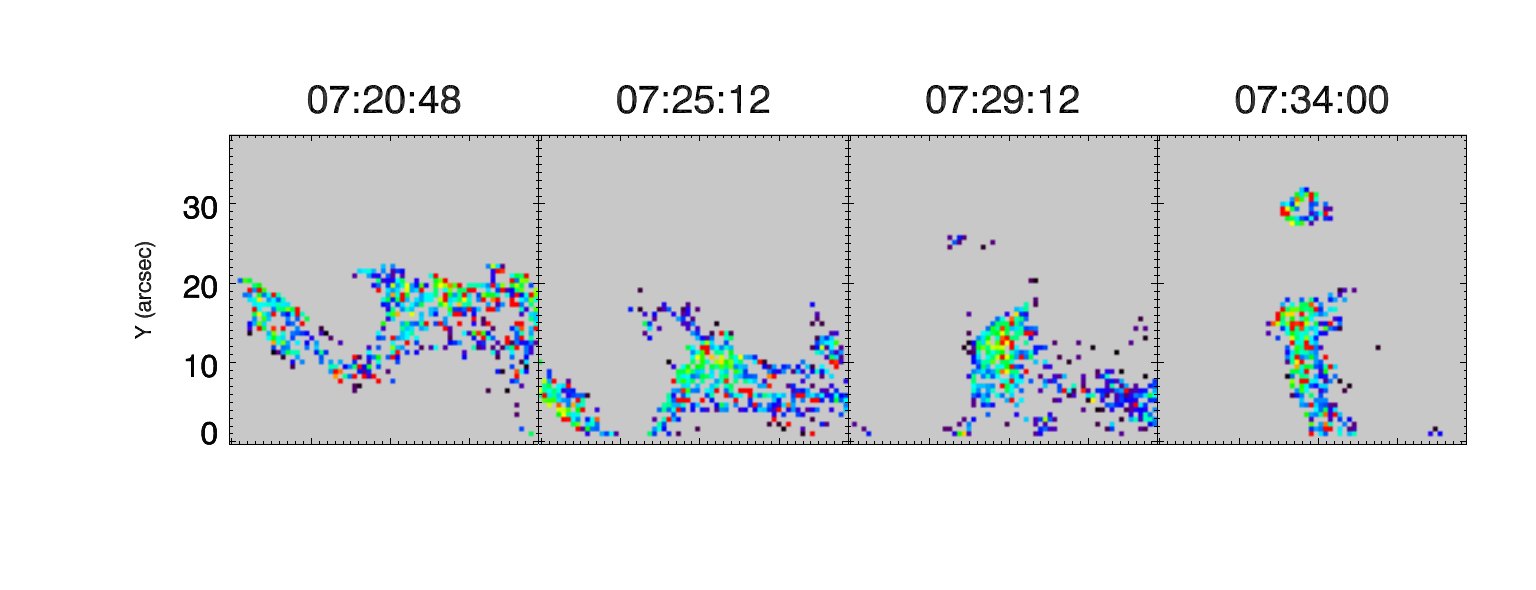} 
\includegraphics[height=5.0cm,trim=69 30 50 10,angle=0]{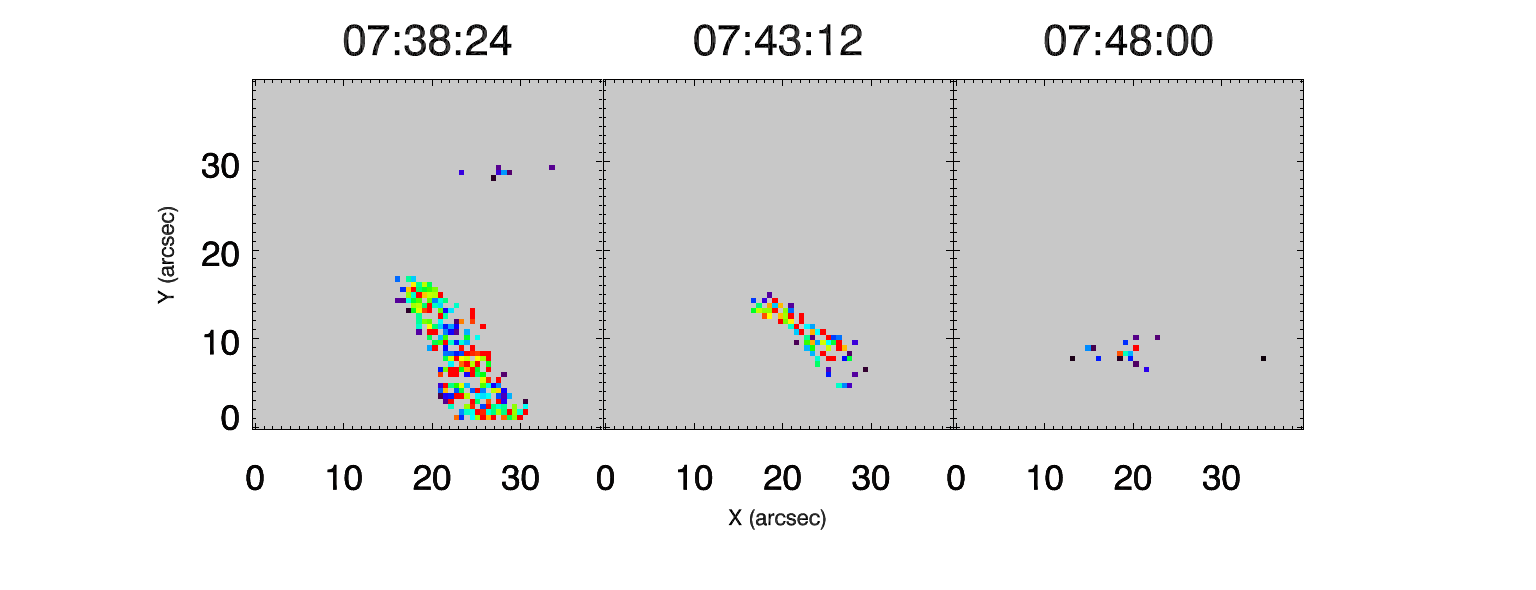} 
\includegraphics[height=4.2cm,trim=0 0 0 50,angle=0]{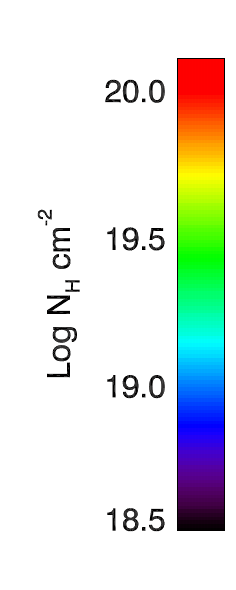}
\caption{Evolution of column density of the blob highlighted in Figure~\ref{snapb} as it passes across the solar disc. The filamentary structure indicative of the magnetic Rayleigh-Taylor instability can be clearly seen. Direction of travel is to the north-east.}
\label{b1}
\end{figure*}

\begin{figure*}
\centering
\includegraphics[height=4.7cm,trim=60 20 160 10,angle=0]{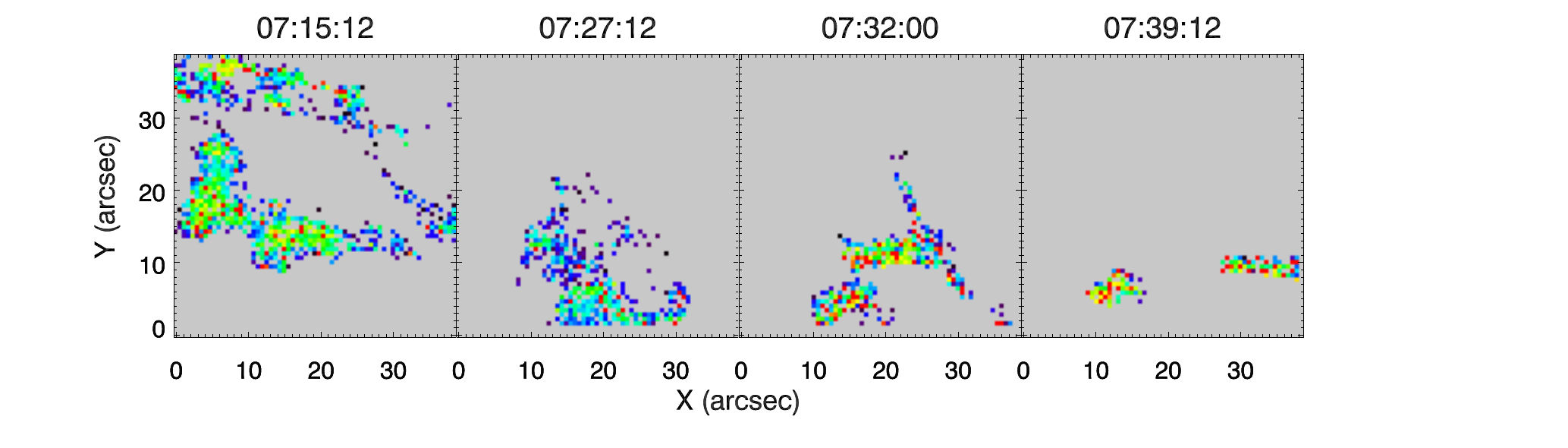}
\includegraphics[width=0.1\textwidth,clip=,trim=10 0 0 0,angle=0]{nh-colourbar-eps-converted-to.pdf}
\caption{Evolution of the density of blob presented in Figures \ref{171} and \ref{A1}. The blob appears to reduce in size but retains a high column density throughout the descent. Direction of travel is to the north-east.}
\label{a25}
\end{figure*}

In order to highlight the time evolution of $N_H$ and $G$, the column density maps of the blob highlighted in Figure~\ref{snapb} between 07:03 and 07:50 UT (when the descent is seen against the solar disc) are presented in Figure~\ref{b1}. Further snapshots of the blob in Figures~\ref{171} and~\ref{A1} are shown in Figure~\ref{a25}. Note that column density, $N_H$, is presented on a logarithmic scale ({\it i.e.}, as log$_{10}(N_H$)). 

\begin{figure*}
\centering
\includegraphics[height=4.7cm,trim=60 20 160 30,angle=0]{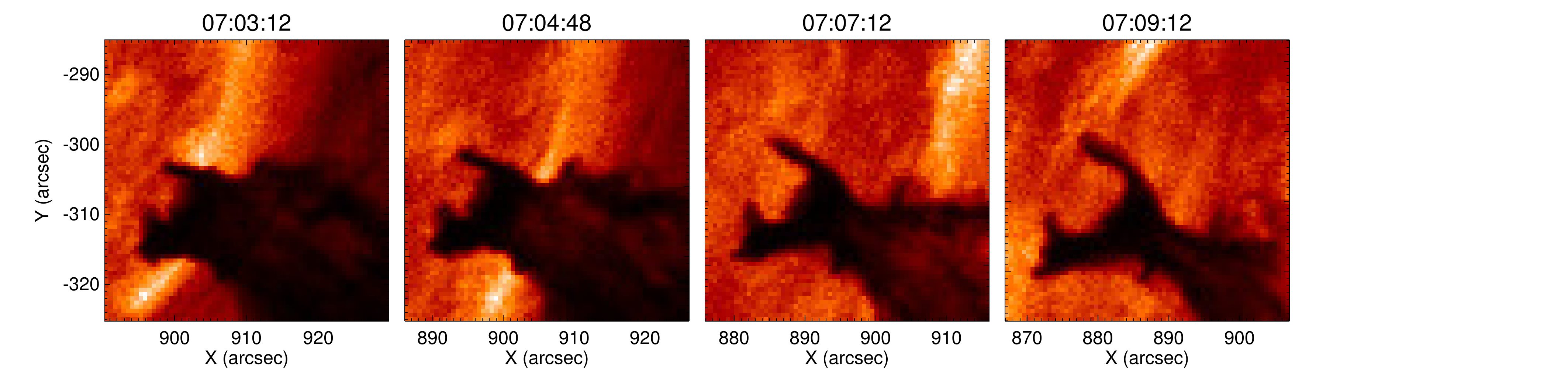}
\caption{Evolution of blob morphology: 171~\AA~images of the first 4 images in Figure~\ref{b1}. Notice the back-end of the blob does not appear in the first N$_H$ map; this is due to dark material lying in the background which leads to little or no difference in intensity between the blob/background images.}
\label{b1-171}
\end{figure*}

\begin{figure*}
\centering
\includegraphics[height=5.0cm,trim=200 20 120 10,angle=0]{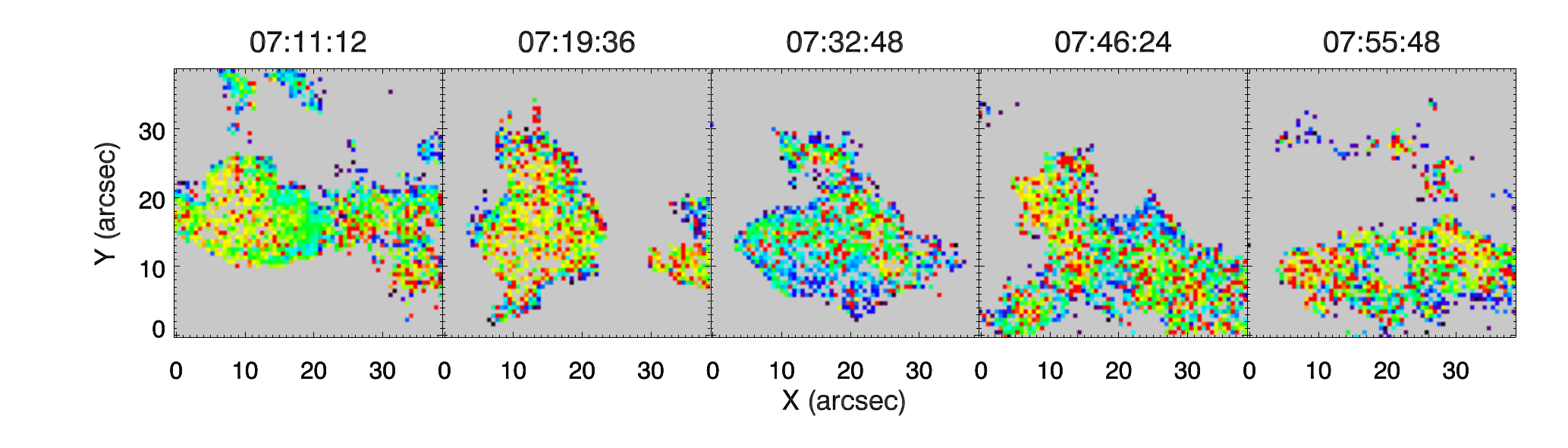}
\caption{Evolution of the density of a third blob (colour bar shown in Figure~\ref{a25} applies). The piling up of the material towards the front of the blob can be seen in the first image and the filamentary structure can be seen forming in the remaining images. Direction of travel is to the north-east.}
\label{blb5}
\end{figure*}

\begin{figure*}
\centering
\includegraphics[height=5.0cm,trim=200 20 120 10,angle=0]{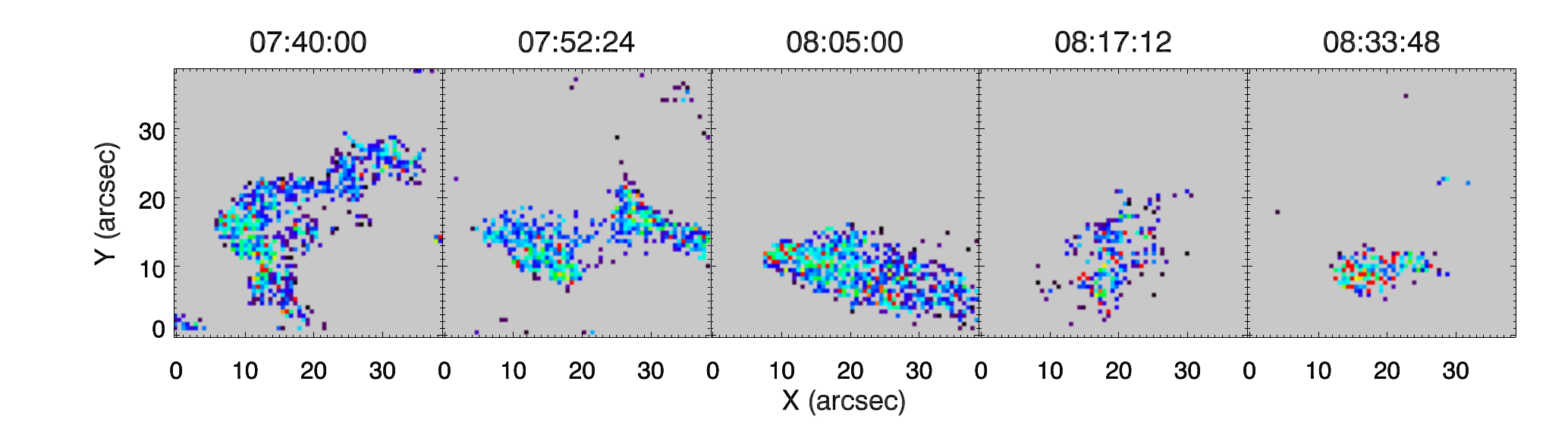}
\caption{Evolution of the density of a fourth blob (colour bar shown in Figure~\ref{a25} applies). The mRT instability is not observed as clearly here, and this blob appears to have a lower column density in general. Direction of travel is to the north-east-east.}
\label{blb5}
\end{figure*}

The figures in this work do not include any pixels where $\chi^2_\nu$~$<$~0.1 or $\chi^2_\nu$~$>$~10, an order of magnitude within a perfect fit. This masking removes pixels with large measurement errors in absorption depth, many of these being outside the visual edge of the blob which still appeared to show $G$~$>$~0.5.

\section{Analysis}

The contours displayed in Figure~\ref{171cont} demonstrate the agreement between the visual edge of the blob and the boundary where $G$ = 0.5. However, it can be seen both from the contours and from comparison with Figure~\ref{A1} that there is a discrepancy towards the tail-end of the blob; whilst there appears to be dark material at approximately (948\arcsec ,-203\arcsec), $G$ here is below 0.5, and as such is not defined as part of the blob. This is due to the fact that the background image contains dark material in this location, which means there is little difference in intensity between this image and the blob image at this point; in other words, the image selected to act as the unobscured background may not be a true reflection of the background radiation field at this point. This is the most suitable co-spatial background image available, but though the observed blobs have well-defined leading edges ({\it i.e.}, towards the direction of travel), most of them have long, diffuse tails, often appearing to be still connected to the original erupted cloud. Therefore, the front edge was attributed greater importance in background-image selection.

Hydrogen column densities were calculated to be approximately 2~$\times$~10$^{19}$~cm$^{-2}$ for the entire descent of all blobs. All blobs appear to decrease in size as time progresses -- {\it i.e.}, as they fall -- but as can be seen from Figures~\ref{b1} and~\ref{a25}, column density values remain reasonably constant. However, the geometric  depth $G$ appears to gradually decrease across the descent from $\sim$0.95 to $\sim$0.8. Although this reduction may also be due to the pixel filling factor falling, the more likely cause is that a greater proportion of the emission measured originates in the foreground as the height of the blobs decreases.

Based on an average value of $N_H$~=~2~$\times$~10$^{19}$~cm$^{-2}$, for a blob of dimensions 10''~x~20'' (the average size of the blobs towards the start of measurement), the initial mass of each blob is approximately 3~$\times$~10$^{13}$~g. By comparing Figures~\ref{snapb}~and~\ref{stereo}, it would seem that the material is elongated roughly in the radial direction, with similar geometries appear in the projections of the blobs seen by both SDO and STEREO. Unfortunately, with observations from only two directions, we are unable to resolve the volumes and positions precisely, and it is possible that our calculations may include multiple blobs lying along the line of sight -- however, the consistency of $N_H$ values derived from AIA images suggests that this is not the case, and the volumetric density of hydrogen for all blobs has been estimated to be of the order 10$^{10}$~cm$^{-3}$. 

The blob used to investigate whether the mRT instability is at work is presented in Figure~\ref{b1}, with the images taken at 07:11 and 07:25 UT used to measure the separation scale between the fingers of material that coalesce into blobs. We estimate this scale to be $\lambda$~$\approx$~1~$\times$~10$^9$~cm. Figure~\ref{stereo} indicates that the width of the blobs appears similar in different projections (STEREO-A, SDO), suggesting that the assumption of cylindrical symmetry is a good approximation. We therefore estimate a volumetric mass density in this blob to be $\rho_u$~=~$\sim$2~$\times$~10$^{-14}$~g~cm$^{-3}$. Using these values of $\lambda$ and $\rho$ in Equation~\ref{domRT2}, with a coronal density of $\sim$10$^{-16}$~g~cm$^{-3}$ (although the result is not sensitive to this value as it is orders of magnitude smaller than the blob density) and $g$~=~2.74~$\times$~10$^4$~cm~s$^{-2}$, and assuming that cos$\theta$~=~1, a magnetic field strength of the order 1~G is expected to be associated with the instability. 

Figure~\ref{b1} also demonstrates the change in density distribution of this blob as it falls through the solar atmosphere. The mass appears to collect, or `pile up' towards the leading edge at certain points, most noticeably just before `forking', splitting into two branches which eventually break into new blobs in their own right; the whole process then repeats. These dynamics are similar to the mRT instability in terms of mass distribution and flow, and indicates that the two regimes of different density ({\it i.e.}, the cool blob and surrounding corona) are being accelerated against one another -- a morphological similarity to the mRT instability.

\section{Discussion}

The values obtained for the hydrogen column density of the blobs analysed in this work are comparable to those found by \cite{Gilbert:2004vs}, who calculated the column density of a prominence to be 1.6~$\pm$~1.0~$\times$~10$^{19}$~cm$^{-2}$ {\it before} eruption -- even after the material in this study was seen to expand greatly in the initial eruption.

In work carried out on the dynamics of this erupted material by \cite{Innes:2012cu}, the instability occurring in the blobs studied in this work was unidentified. However the mRT was seen to be at work in material from the same eruption higher in the corona, which was corroborated by a realistic Alfv\'en velocity inferred from the separation of the forks. Our finding that the density evolution is self-similar in all the blobs studied strongly implies the occurrence of an instability, and the morphological similarity between these blobs and the work carried out by \cite{Innes:2012cu} points towards the mRT instability.

It is important to note that our results are in fact a lower limit on column density, since they depend on bound-free absorption by hydrogen or helium \citep{Williams:2013bq}. We also note that as the material falls, the blobs appear to shrink, suggesting a depletion of bound electrons (ionisation), perhaps by heating. Nonetheless, the fact that the column densities of the blobs remain reasonably constant in the interior is an intriguing result, and suggests that the cool material is somewhat isolated from the surrounding environment. Given the relatively high density of cool material, thermal conduction into the interior is likely to be inefficient. 

This method only considers differences in intensity arising from absorption against a model background radiation field. However, the possibility that emission could be occurring in the blobs is a source of uncertainty in the results, and this could be a factor influencing the goodness of the fit. However, it is unlikely to be a profound effect, and as the majority of the pixels have low $\chi^2_\nu$ values (below 10), the drop in intensity caused by the passing material seems in good agreement with photo-absorption by low-temperature material.

The morphology and dynamics of the blobs are not only self-similar, but are more specifically indicative of the mRT instability. When we assume this is the case, the calculated density values and length scales observed in the investigation lead us to infer a magnetic field strength of order 1~G associated with the material. This seems to be reasonable: values of 0.4~--~1.3~G at a height of 1.6~--~2.1~R$_\odot$ were found by \cite{Cho:2007fq}, and active region prominences have been shown to have magnetic fields up to 100~G, which could feasibly become 1~G within these blobs following the huge lateral expansion that this material undergoes during eruption. However, Equation~\ref{domRT} only considers structuring formed along the magnetic field (constrained by tension along the field lines).

Further to this, should the magnetic field be purely in one direction, we would expect there to be no suppression of small modes across this direction; we might therefore expect to see the front of the blobs decreasing in column density as the nonmagnetic RT instability occurs in a direction which we are unable to observe. We notice this, for example in the 07:32 image of Figure~\ref{blb5}, but this is not generally seen across the blobs. Therefore, a suppression of the smallest modes in all horizontal directions is necessary. This is where shear between the magnetic fields (as found in the simulations of \citealp{2007ApJ...671.1726S} and \citealp{2012ApJ...746..120H}) can become important. The presence of magnetic shear means that there is no direction that is perpendicular to both the magnetic field in the blob and in the corona (whilst also perpendicular to the interface), increasing the size of the smallest scales that can be formed. If we assume that the result $B$~=~1~G gives the strength of the sheared component of the magnetic field, then this will be a lower limit for the coronal/blob field strength. It should also be noted that the estimate for $\lambda$ used to calculate $|B|$ is also a lower limit, which would also result in an underestimate of the field strength.

\section{Conclusion}

After an exceptionally large amount of material was thrown into the solar atmosphere by a filament eruption on the 2011~June~7, discrete blobs of plasma were seen to fall back to the solar surface. We have calculated the column density of the in-falling blobs to be approximately 2 $\times$ 10$^{19}$ cm$^{-2}$, which is comparable with {\it pre-eruption} column densities of filaments.

We have studied the evolution of the blobs as they fall through the solar atmosphere and find morphological and dynamic evidence that they are formed by the magnetic Rayleigh-Taylor instability. The shapes that the plasma takes are similar to those seen in simulations of this instability, and the dynamics of the density distribution within the blobs further support this. We therefore conclude that the returning blobs from this enormous CME appear to undergo the magnetic Rayleigh-Taylor instability.

By using the point of view given by STEREO-A, we can reasonably approximate the geometry of the blobs as cylindrical, which leads to an approximate volumetric density of 2~$\times$~10$^{-14}$~g~cm$^{-3}$. By measuring a separation of 10$^9$~cm in the forking material, we then use Equation~\ref{domRT2} to infer a magnetic field strength associated with the mRT instability of the order 1~G.

\acknowledgements{J.C. thanks UCL and MPI for an Impact PhD Studentship. The research leading to these results has received funding from the European Commission's Seventh Framework Programme under the grant agreement No. 284461 (eHEROES project). L.v.D.G.'s work was supported by the STFC Consolidated Grant ST/H00260X/1 and the Hungarian research grant OTKA K-081421. A.H. is supported by KAKENHI Grant-in-Aid for Young Scientists (B) 25800108.}

\end{document}